\documentclass[aps,prl,reprint,superscriptaddress]{revtex4-2}

\usepackage{amsmath,amssymb,graphicx,bm}
\usepackage[hidelinks]{hyperref}

\begin{document}

\title{From Berg--Purcell precision bounds to clock-limited information capacity}

\author{Micha\l\  Komorowski}
\email{m.komorowski@sysbiosig.org}
\affiliation{Institute of Fundamental Technological Research, Warsaw, Poland}


\begin{abstract}
Physical limits to chemical sensing are traditionally expressed as Berg-Purcell bounds on estimation accuracy. 
Whether these bounds also limit the total amount of information a molecular receptor can transmit has remained unclear, despite the fact that cellular signaling performance is naturally quantified in bits rather than precision alone.
Here we derive an explicit link between Berg-Purcell-type sensing limits and the information capacity of a single two-state receptor, yielding a compact expression that separates contributions from concentration range, receptor copy number, and averaging time.
We show that, in the ideal fixed-time occupancy model, diffusion-limited sampling alone does not define a finite global information bound: with a perfectly specified integration window and unbounded input range, information capacity grows without bound with dynamic range, albeit slowly.
A finite saturation arises when time integration is treated as an explicit physical resource.
Finite timing precision yields a clock-limited bound on information capacity, and in the high-occupancy regime information transmission crosses over from diffusion-limited to clock-limited behavior.
Together, these results establish a receptor-level information bound in bits that is finite once constraints on timing precision are taken into account.
\end{abstract}

\maketitle

Physical limits to chemical sensing were established in the classic work of Berg and Purcell~\cite{BergPurcell1977}, 
who related the uncertainty of concentration measurements to the diffusive renewal of molecules and stochastic receptor occupancy. 
Subsequent analyses refined this picture and yielded bounds on concentration–estimation error, typically expressed as a fractional uncertainty $(\delta c/c)^2$ after an averaging time $T$~\cite{BialekSetayeshgar2005,Mora2015,tenWolde2016,EndresWingreen2009,Kaizu2014,Aquino2016,HandyLawley2021}. 
These results form the foundation of modern theories of biochemical sensing.

At the same time, contemporary single-cell experiments increasingly quantify sensing fidelity in informational terms, measuring mutual information and information capacity in bits~\cite{Shannon1948,CoverThomas2006,Cheong2011,Selimkhanov2014,TkacikTenWolde2025}. 
From this perspective, the natural question is not only how precisely a concentration can be estimated, but how many distinct concentration levels can be reliably discriminated over a finite dynamic range. 
Classical Berg--Purcell--type bounds do not directly answer this question.
Berg--Purcell theory addresses local estimation precision at a specified concentration and observation time, whereas information capacity is a global measure that also depends on the allowed input ensemble.

To address this gap, one must translate local estimation precision into a global, information-theoretic measure~\cite{Jeffreys1946,ClarkeBarron1990,BrunelNadal1998,Jetka2018,Komorowski2025}. 
Here we provide such a translation in closed form for the canonical single-receptor, two-state binding model. 
When expressed in information-theoretic terms, classical precision bounds reveal a dependence on global input constraints: under the idealized assumption of a perfectly defined integration window and an unbounded input range, diffusion-limited sampling alone does not define a finite information capacity. 
Instead, capacity grows without bound with input range, albeit only log-logarithmically.

This translation is physically revealing but also highlights severe diminishing returns. 
For a single receptor, gaining an additional bit of information requires an approximately fourfold increase in observation time or receptor copy number, making multi-bit transmission difficult without parallelization.  
A finite saturation of information transmission arises in the model developed here when time integration itself is treated as a constrained physical resource. 
Accounting for finite timing precision therefore yields a clock-limited information bound for single-receptor sensing within this fixed-window model. Although motivated by biochemical receptors, the underlying argument is not system-specific and applies generally to physical sensors that infer signals from stochastic counting processes under finite time resolution.

Following the classical modeling framework~\cite{BergPurcell1977,BialekSetayeshgar2005,Mora2015,tenWolde2016}, we consider a single receptor molecule that switches between an unbound state $U$ and a bound state $B$,
\begin{equation}
U\underset{k_-}{\overset{k_+ c}{\rightleftharpoons}}B,
\end{equation}
where $c$ is ligand concentration, $k_+$ the association rate constant, and $k_-$ the dissociation rate.
The steady-state occupancy is
\begin{equation}
p(c)=\frac{c}{c+K_d},\qquad K_d=\frac{k_-}{k_+}.
\end{equation}
Over an observation window of duration $T$, the receptor generates a stochastic trajectory (Figure 1); canonical readouts include
the fraction of time bound $Y=T_B/T$~\cite{BergPurcell1977,BialekSetayeshgar2005,tenWolde2016}, the count of binding/unbinding events,
or (in principle) the full event-time series used in maximum-likelihood estimation~\cite{EndresWingreen2009,Mora2015,tenWolde2016}. Here we focus on Berg–Purcell sensing based on the time-averaged occupancy $Y=T_B/T$~\cite{BergPurcell1977,tenWolde2016}.
In diffusion-limited binding, $k_+$ is determined by geometry and diffusion~\cite{BergPurcell1977,BialekSetayeshgar2005}. Figure~1 highlights the implicit timing assumption behind idealized Berg–Purcell sensing and motivates treating time integration as a constrained physical resource~\cite{tenWolde2016}.

Let $P(Y|c)$ denote the conditional distribution of a chosen readout $Y$ given concentration $c$.
The Fisher information (FI) about $c$ is
\begin{equation}
\mathcal{I}(c)=\left\langle\left(\frac{\partial}{\partial c}\ln P(Y|c)\right)^2\right\rangle.
\end{equation}
For a large effective sample size, i.e., large receptor copy number, $N$, efficient estimators approach the Cram\'er-Rao bound \cite{Fisher1925,Rao1945,Cramer1946,LeCam2012}
$\mathrm{Var}(\hat c)\simeq \mathcal{I}(c)^{-1}$. To convert local distinguishability (FI) into a global, distribution-free fidelity measure,
we use the asymptotic capacity expression for one-dimensional inputs~\cite{ClarkeBarron1990,Jeffreys1946,Jetka2018,Komorowski2025}:
\begin{equation}
C_A^\ast \;=\; \log_2\!\left(\frac{1}{\sqrt{2\pi e}}\int_{c_{\min}}^{c_{\max}}\!\sqrt{\mathcal{I}(c)}\,dc\right).
\label{eq:asymptotic_capacity}
\end{equation}
As discussed in Supplementary Sec.~S1, Eq.~\eqref{eq:asymptotic_capacity} approximates the Shannon information capacity in the large-sample (large-$N$) regime. Precisely, $C_A^\ast$ is the asymptotic capacity, introduced for biochemical signaling in Refs.~\cite{Jetka2018,Komorowski2025}: for $N$ independent replicas of the same channel (i.e., $N$ independent receptors), the Shannon capacity satisfies $C_N^\ast\xrightarrow{} C_A^\ast+\tfrac{1}{2}\log_2 N$ with increasing $N$, so $C_A^\ast$ fixes the $N$-independent offset and thus determines the number of distinguishable inputs for a given system with $N$ receivers. Related bounds for collective concentration sensing in communicating cell populations have been derived in Ref.~\cite{FancherMugler2017}. Because it is an additive constant (offset) in this scaling, $C_A^\ast$ can take negative values, which does not contradict nonnegativity of the exact Shannon capacity~\cite{Jetka2018,Komorowski2025}.
Assuming large copy number, $N$, of identical receptors observed in parallel, we write
\begin{equation}
C_N^\ast \;=\; C_A^\ast + \frac{1}{2}\log_2 N.
\label{eq:N_scaling}
\end{equation}

In the Berg–Purcell framework, the cell estimates concentration from the time-averaged occupancy $Y=T_B/T$ over a fixed integration window $T$. For long $T$ this yields the fractional estimation error
\begin{equation}
\left(\frac{\delta c}{c}\right)^2 \simeq \frac{2}{k_-T}\,\frac{1}{p(c)},
\label{eq:bp_variance}
\end{equation}
Using $p(c)=c/(c+K_d)$, Eq.~\eqref{eq:bp_variance} implies FI (Supplementary Sec.~S2)
\begin{equation}
\mathcal{I}(c)
\;\simeq\;
\frac{k_-T}{2}\,\frac{1}{c(c+K_d)}.
\label{eq:FI_ideal}
\end{equation}
Substituting Eq.~\eqref{eq:FI_ideal} into Eq.~\eqref{eq:asymptotic_capacity} and assuming $c_{\min}\!\ll\!K_d$ and $c_{\max}\!\gg\!K_d$ yields a closed-form integral (Supplementary Sec.~S3):
\begin{equation}
C_N^\ast
=
\log_2\!\left[
\frac{2}{\sqrt{2\pi e}}\,
\ln\!\left(2\sqrt{\frac{c_{\max}}{K_d}}\right)
\right]
+
\frac{1}{2}\log_2\!\left(\frac{k_-TN}{2}\right).
\label{eq:capacity_asymptotic}
\end{equation}

Eq.~\eqref{eq:capacity_asymptotic} separates cleanly into:
(i) a \emph{baseline (dynamic-range) term} that depends only on $c_{\max}/K_d$,
(ii) a \emph{temporal averaging term} $\tfrac{1}{2}\log_2(k_-T)$ set by the number of independent switching opportunities,
and (iii) a \emph{copy-number term} $\tfrac{1}{2}\log_2 N$ from parallel independent receptors. Doubling $T$ or $N$ increases capacity by $0.5$ bits. 

Equation~\eqref{eq:capacity_asymptotic} implies three general consequences.
\emph{(1) Strong diminishing returns of dynamic range.}
Capacity grows with dynamic range only as $\log_2\!\ln(\sqrt{c_{\max}/K_d})$. Increasing the largest resolvable concentration therefore yields little additional information compared to increasing the number of independent samples (larger $T$) or receptors ($N$).
\emph{(2) A “half-bit per doubling” law for time and receptor number.}
Because the sampling contribution enters as $\tfrac12\log_2(k_-TN)$, doubling $T$ or $N$ increases the bound by approximately $0.5$ bits. Achieving an additional bit thus requires roughly a fourfold increase in averaging time or receptor copy number.
\emph{(3) The ideal fixed-time capacity depends on the upper input limit.}
Increasing dynamic range yields only diminishing returns:
doubling $c_{\max}$ changes the baseline only through $\ln(\sqrt{c_{\max}})$. Equation~\eqref{eq:capacity_asymptotic} implies that, in the ideal fixed-time occupancy model, the information capacity grows without bound as $c_{\max}\to\infty$, albeit extremely slowly: $C_N^\ast \sim \log_2 \ln c_{\max}$. It follows from combining local relative precision with an increasingly large input domain and a perfectly specified integration window. At high concentration, Eq.~\eqref{eq:FI_ideal} gives $\mathcal{I}(c)\sim (k_-T/2)\,c^{-2}$, implying constant precision in $\ln c$ as the input range is extended (see Supplementary Sec.~S3.1).
Thus Eq.~\eqref{eq:capacity_asymptotic} should be read as the capacity of an idealized fixed-time model over a specified input interval, not as a failure of diffusion-limited sensing theory. Diffusive sampling alone does not define a finite global information bound without additional constraints on input support or measurement resources. In the next step we introduce finite timing precision as one minimal physical constraint that regularizes this idealization.

We therefore introduce finite timing precision as a physically motivated completion of the ideal fixed-time model by treating the downstream averaging window as stochastic. This is one minimal regularization of the global capacity problem.
The downstream circuit does not know the averaging window $T$ perfectly.
Precisely, we keep the bound time $T_B$ accumulated over the intended window $T$, but allow the downstream normalization to use an imperfect internal duration estimate $T_v$, with mean $\langle T_v\rangle=T$ and variance $\sigma_T^2$.
The resulting occupancy-like readout is $Y=T_B/T_v$ (Supplementary Sec.~S4.1).
This 'noisy clock' model can be interpreted as finite precision in reading out the elapsed averaging time, rather than as a stochastic stopping time for the receptor trajectory
(e.g.\ due to stochastic molecular turnover in a timer network).

Propagating this uncertainty adds a contribution to the fractional concentration error (Supplementary Sec.~S4.2):
\begin{equation}
\begin{aligned}
\left(\frac{\delta c}{c}\right)^2
&\simeq
\underbrace{\frac{2}{k_-T}\,\frac{1}{p(c)}}_{\text{sensing noise / occupancy averaging}}\\
&\quad+
\underbrace{\frac{\sigma_T^2}{T^2}\left(\frac{1}{1-p(c)}\right)^2}_{\text{clock noise}},
\end{aligned}
\label{eq:relerr_clock}
\end{equation}
Because timing uncertainty enters through the random denominator $T_v$, it adds a multiplicative contribution to the relative concentration error: this contribution approaches $(\sigma_T/T)^2$ as $p\to0$, but becomes dominant only at high occupancy, where it changes the Fisher-information scaling from $c^{-2}$ to $c^{-4}$, and thereby adds to the inverse Fisher information (Supplementary Sec.~S4.3),
\begin{equation}
\mathcal{I}_{\mathrm{clock}}(c)^{-1}
\simeq
\frac{2}{k_-T}\,c(c+K_d)
+
\frac{\sigma_T^2}{T^2}\,\frac{c^2(c+K_d)^2}{K_d^2}.
\label{eq:FI_clock}
\end{equation}

Equation~\eqref{eq:FI_clock}
makes explicit that clock noise reduces Fisher information at all concentrations, placing the ideal-timing result as a strict upper bound, $\mathcal{I}_{\mathrm{clock}}(c)\le \mathcal{I}(c)$ pointwise in $c$, so the clock-imprecise capacity is always below the ideal-timing bound. Moreover, the capacity integral can still be evaluated in closed form.
Let $x\equiv c/K_d$, $A\equiv 2/(k_-T)$, and $s^2\equiv (\sigma_T/T)^2$.
Then (Supplementary Sec.~S4.4)
\begin{equation}
\int_{c_{\min}}^{c_{\max}}\!\sqrt{\mathcal{I}_{\mathrm{clock}}(c)}\,dc
=
\sqrt{\frac{k_-T}{2}}
\Big[\mathrm{F}(\phi_{\max}\,|\,\mu)-\mathrm{F}(\phi_{\min}\,|\,\mu)\Big],
\label{eq:V_clock_closed}
\end{equation}
where $\mathrm{F}(\cdot|\mu)$ is the incomplete elliptic integral of the first kind,
\begin{equation}
\begin{aligned}
\sin\phi(c)&=\frac{2\sqrt{x(1+x)}}{1+2x},\\
\mu&=1-\frac{s^2 k_-T}{8}
=1-\frac{\sigma_T^2}{T^2}\,\frac{k_-T}{8}.
\end{aligned}
\end{equation}
Consequently,
\vspace{0.2cm}
\begin{equation}
\begin{aligned}
C_{\mathrm{clock}}^{\ast}
&=
\log_2\!\Bigg[
\frac{1}{\sqrt{2\pi e}}
\sqrt{\frac{k_-TN}{2}}\\
&\qquad\times
\Big(\mathrm{F}(\phi_{\max}|\mu)-\mathrm{F}(\phi_{\min}|\mu)\Big)
\Bigg],
\end{aligned}
\label{eq:capacity_clock_closed}
\vspace{0.2cm}
\end{equation}
 where the $\tfrac{1}{2}\log_2 N$ gain holds for $N$ independent receptors with independent clock noise.
 Equation~\ref{eq:capacity_clock_closed} thus gives a clock-regularized receptor-level information bound; its dependence on system parameters is made explicit through the function $F(\phi|\mu)$, evaluated between the physical limits $\phi_{\min}$ and $\phi_{\max}$.

Crucially, for $c\gg K_d$, according to Eq.~\eqref{eq:FI_clock},
$\mathcal{I}_{\mathrm{clock}}(c)\sim \frac{K_d^2T^2}{\sigma_T^2\,c^4}\qquad (c\gg K_d)$, so $\int^\infty_0 \sqrt{\mathcal{I}_{\mathrm{clock}}(c)}\,dc$ converges and the capacity remains finite even as $c_{\max}\to\infty$. 
Finite timing precision therefore provides one explicit mechanism for saturation of information capacity: in the high-occupancy regime of this model, information transmission becomes clock-limited rather than diffusion-limited.
A convenient microscopic model of a noisy clock is an $m$-step exponential timer:
$T_v=\sum_{i=1}^m v_i$ with $v_i$ i.i.d.\ exponential (Erlang clock).

This construction yields a coefficient of variation $\sigma_T/T = 1/\sqrt{m}$, controlled solely by the number of clock steps $m$. In this interpretation, $m$ quantifies the biochemical resources invested in temporal precision. Figure~1 schematically illustrates how such a stochastic clock provides an internal estimate of the elapsed duration used to normalize the occupancy signal.

The consequences of finite timing precision for information transmission are shown in Figures~2 and~3. Figure~2 demonstrates that, once timing precision is finite, increasing the mean integration time $T$ beyond a clock-dependent scale yields even stronger diminishing returns: the ideal $\tfrac{1}{2}\log_2 T$ growth predicted by Berg–Purcell scaling no longer holds and the capacity approaches a plateau. Figure~3 shows an analogous saturation with concentration dynamic range. While ideal timing implies a slow but unbounded growth $C_{N}^{\ast} \sim \log_2\!\ln c_{\max}$, finite clock precision changes the high-concentration scaling of the Fisher information from $c^{-2}$ to $c^{-4}$, rendering the capacity integral convergent. In this regime, information transmission becomes clock-limited rather than diffusion-limited. Equation~(13) therefore provides a closed-form receptor-level information bound in bits for this timing-regularized model.

The central conceptual message of this work is that diffusion-limited receptor sampling alone does not define a finite global information bound without additional constraints on input range or measurement resources. Berg--Purcell theory constrains local concentration-estimation precision, but this local precision bound does not by itself determine how many distinct input states can be reliably distinguished over a specified concentration domain. In the ideal fixed-time model, extending that domain yields a weak but unbounded $\log_2\!\ln c_{\max}$ growth.

Several physical constraints could regularize this global capacity problem, including finite input support, receptor copy number variation, downstream noise, or energetic limits. Dynamic-input formulations, in which information is carried by time-dependent input and output trajectories, provide a complementary framework for fluctuating environments~\cite{TostevinTenWolde2009,MoraNemenman2019}. Here we deliberately focus on the static fixed-window occupancy problem because this is the setting in which Berg--Purcell--type precision bounds are most commonly stated, and it provides a controlled way to ask how local precision bounds translate into receptor-level information bounds. When the normalization time itself is uncertain, this uncertainty becomes dominant at high occupancy and produces an analytically transparent crossover from diffusion-limited to clock-limited behavior.

The implied constraints are consistent with empirical estimates of information transmission in biological signaling systems, which frequently lie in the low-single-bit regime. For example, early Drosophila patterning through the Bicoid-Hunchback axis transmits approximately $1.5$-$1.7$ bits, while combined readout of multiple gap genes yields several bits of positional information~\cite{TkacikCallanBialek2008,TkacikEtAl2015PosInfo}. Similarly, it is argued that although receptors may operate close to physical sensing limits \cite{CelaniVergassola2010,BrumleyEtAl2019,MicaliEtAl2017,WanJekely2021}, much of this sensory information is lost in downstream processing, rendering chemotaxis far less efficient than receptor-level bounds would suggest \cite{mattingly2026coli,TuGrinstein2005,MatthaeusEtAl2011}. Besides, in eukaryotic signaling, population-level measurements often yield less than one bit, whereas higher values emerge only through repeated stimulation, multiple effectors, or parallel pathways~\cite{TopolewskiKomorowski2021}. Eq.~\eqref{eq:capacity_clock_closed} shows that such low information values are compatible with receptor-level constraints of this kind, although they should not be read as evidence for timing precision being the limiting factor in any particular system.

More broadly, these results help rationalize why biological systems rarely rely on extreme dynamic range or prolonged averaging at the level of individual receptors. Because capacity scales so weakly with these parameters, expanding signaling capacity through receptor duplication, diversification, and parallel wiring is a far more efficient strategy~\cite{KomorowskiTawfik2019,CarballoPacheco2019}. This provides a physical underpinning for the widespread use of cross-wired architectures and receptor families observed in cellular signaling networks~\cite{antebi2017combinatorial,amit2007evolvable}.

Viewed in this way, classical sensing limits acquire a new interpretation. When Berg--Purcell--type local precision bounds are combined with explicit constraints on the input domain and measurement resources, they can be converted into receptor-level information bounds. Finite timing precision provides one analytically tractable example, yielding a clock-limited bound for the two-state receptor model considered here. Information, alongside precision, therefore provides a natural currency for assessing the capabilities of biochemical signaling systems.

\begin{figure}[t]
  \centering
  \includegraphics[width=0.98\columnwidth]{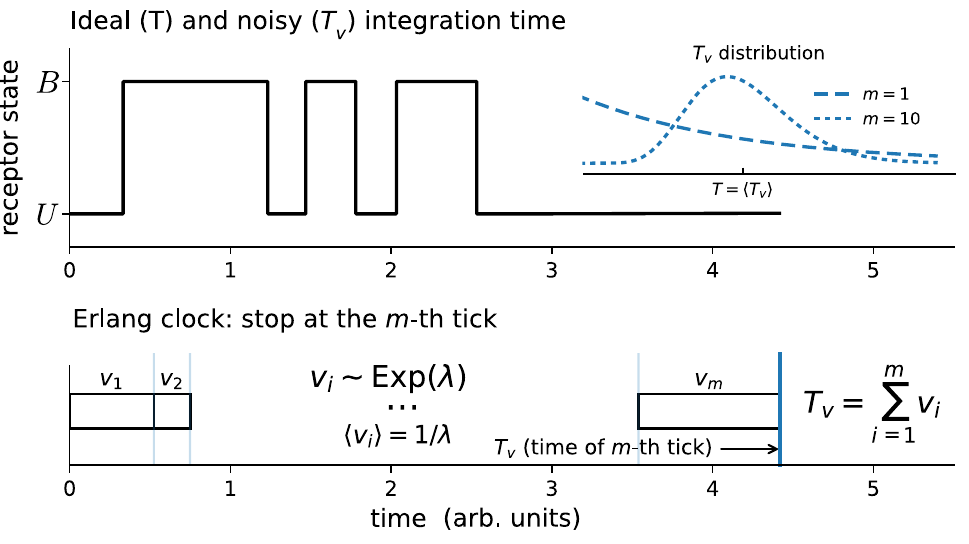}
  \caption{\label{fig:schematic_clock}
Single-receptor trajectory and the physical origin of a clock-limited information bound. Idealized Berg–Purcell sensing assumes the cell forms the time-averaged occupancy $Y=T_B/T$ over a well-defined integration window $T$. This schematic highlights the implicit assumption of a perfectly defined integration window underlying ideal Berg–Purcell sensing.
In the clock-noisy model used here, the bound time $T_B$ is accumulated over the intended window $T$, while a separate biochemical clock provides the noisy duration estimate $T_v$ used for normalization. In the illustrative Erlang clock, $T_v$ is defined as the time of the $m$-th tick of a renewal process, $T_v=\sum_{i=1}^{m} v_i,$
where the inter-tick intervals $v_i$ are independent exponentially distributed random variables with mean $\langle v_i\rangle=1/\lambda$.
Finite timing precision then limits how much information can be extracted from increasingly rapid switching at high ligand concentration.}
\end{figure}

\begin{figure}[t]
  \centering
  \includegraphics[width=0.98\columnwidth]{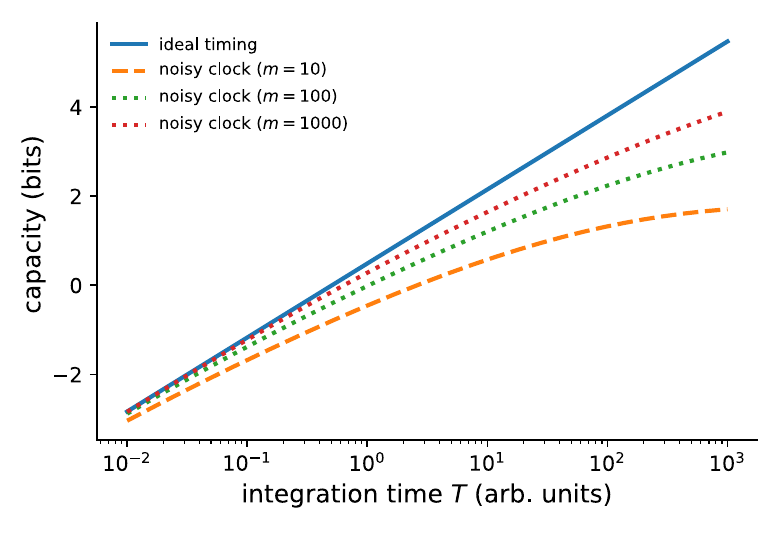}
 \caption{\label{fig:capacity_vs_time}
 Capacity ($C^*_A$) vs mean integration time $T$ for a fixed dynamic range.
Here $C_A^*$ is the additive asymptotic offset in Eq.~\eqref{eq:N_scaling}, plotted for a single receptor ($N=1$), so the full $N$-receptor capacity is $C_N^*=C_A^*+\tfrac12\log_2 N$; negative values indicate the effectively zero-bit single-receptor regime, not negative mutual information.
With ideal time integration (solid), the Berg–Purcell bound predicts unbounded growth of capacity with averaging time (Eq.~\eqref{eq:capacity_asymptotic}).
When time integration is implemented by a stochastic biochemical clock (dashed/dotted), capacity saturates beyond a clock-dependent scale, beyond which longer averaging yields diminishing returns. 
(Parameters:
$K_d = 1$ (concentration unit);
$k_- = 1$ (time unit);
$c_{\min} = 10^{-2} K_d$;
$c_{\max} = 10^{2} K_d$;
$m = 10,\,100,\,1000$.)
}
\end{figure}

\begin{figure}[t]
  \centering
  \includegraphics[width=0.98\columnwidth]{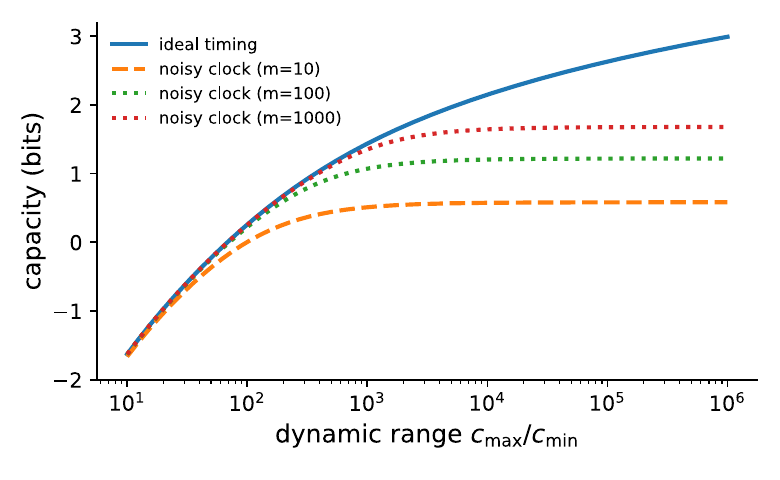}

  \caption{\label{fig:capacity_vs_dynamic_range}
Capacity ($C^*_A$) vs dynamic range $c_{\max}/c_{\min}$ at fixed $T$.
With ideal timing (solid), Berg–Purcell scaling predicts a slow but unbounded growth of capacity, $C^*_A \sim \log_2 \ln c_{\max}$ (Eq.~\eqref{eq:capacity_asymptotic}).
Finite timing precision (dashed/dotted) changes the high-concentration scaling of the Fisher information, causing the capacity to saturate and identifying a clock-limited regime at large dynamic range (Eq.~\eqref{eq:capacity_clock_closed}).
As in Figure~\ref{fig:capacity_vs_time}, $C_A^*$ is the single-receptor ($N=1$) offset.
Parameters as in Figure~\ref{fig:capacity_vs_time}, except: fixed integration time $T = 10$; $c_{\max}/c_{\min} \in [10,10^6]$.
}
\end{figure}

\bibliographystyle{apsrev4-2}
\bibliography{references}

\section*{Acknowledgements}
This research was supported by the National Science Center, Poland, under grant number 2020/39/B/NZ2/03259.

\section*{Author contributions}
M.K. conceived and designed the study, performed the analyses, and wrote the manuscript.

\section*{Competing interests}
The author declares no competing interests.

\clearpage
\onecolumngrid
\setcounter{equation}{0}
\renewcommand{\theequation}{S\arabic{equation}}
\setcounter{figure}{0}
\renewcommand{\thefigure}{S\arabic{figure}}
\setcounter{table}{0}
\renewcommand{\thetable}{S\arabic{table}}
\begin{center}
{\large\bfseries Supplemental Material}\\[4pt]
{\itshape From Berg--Purcell precision bounds to clock-limited information capacity}
\end{center}
\medskip

This Supplemental Material provides the technical details and full derivations underlying the analytical results presented in the main text, in order to make all calculations  explicit and to clarify assumptions and approximations that are stated only briefly in the main paper. Throughout we consider the canonical two-state receptor \cite{BergPurcell1977,EndresWingreen2009},
\begin{equation}
U\underset{k_-}{\overset{k_+ c}{\rightleftharpoons}}B,
\end{equation}
with ligand concentration $c$, association constant $k_+$, dissociation rate $k_-$, and $K_d\equiv k_-/k_+$. The equilibrium occupancy is
\begin{equation}
 p(c)=\frac{c}{c+K_d},\qquad 1-p(c)=\frac{K_d}{c+K_d}.
\end{equation}

\section*{S1. From FI to asymptotic capacity}

\subsection*{S1.1 Asymptotic capacity from FI}

For a one-dimensional input parameter $c\in[c_{\min},c_{\max}]$ and output $Y$ drawn from $P(Y\,|\,c)$, the Fisher information is
\begin{equation}
\mathcal I(c)=\Big\langle\Big(\partial_c\ln P(Y\,|\,c)\Big)^2\Big\rangle.
\end{equation}

The FI-based asymptotic capacity used in the main text is (see Ref.~\cite{Jetka2018,Komorowski2025} for a derivation and conditions)
\begin{equation}
C_A^* = \log_2\!\left(\frac{1}{\sqrt{2\pi e}}\int_{c_{\min}}^{c_{\max}}\!\sqrt{\mathcal I(c)}\,dc\right).
\label{eq:SI_cap}
\end{equation}

The quantity $C_A^*$ is an asymptotic approximation to Shannon capacity: it captures the $N$-independent offset in the large-$N$ scaling $C_N^*\approx C_A^*+\tfrac12\log_2 N$ for $N$ independent replicas \cite{Jetka2018,Komorowski2025}. Because it is defined as this offset, $C_A^*$ can take negative values; negative $C_A^*$ simply indicates  low  number of resolvable inputs for high $N$ and is compatible with the exact Shannon capacity being nonnegative (approaching $0$ in the low-information regime)~\cite{Jetka2018,Komorowski2025}.

A compact way to motivate Eq.~\eqref{eq:SI_cap} is via the large-sample expansion of mutual information in Bayesian estimation.
For $M$ independent samples $Y_1,\ldots,Y_M$ drawn from $P(Y\,|\,c)$ under a prior $\pi(c)$, the mutual information satisfies
\begin{equation}
I(c;Y_1,\ldots,Y_M) = H(\pi) - \big\langle H\big(P(c\,|\,Y_1^M)\big)\big\rangle,
\end{equation}
where $H$ is differential entropy. Under standard regularity conditions, the posterior becomes asymptotically normal with variance $[M\mathcal I(c)]^{-1}$ (Bernstein--von Mises theorem), giving
\begin{equation}
I(c;Y_1^M)= \frac{1}{2}\log_2\!\left(\frac{M}{2\pi e}\right) + \int \pi(c)\,\log_2\!\left(\sqrt{\mathcal I(c)}\right)dc - H(\pi) + o(1).
\end{equation}
Maximizing over priors yields the Jeffreys prior $\pi^*(c)\propto\sqrt{\mathcal I(c)}$ and the resulting maximum mutual information is Eq.~\eqref{eq:SI_cap} (with $\mathcal I$ understood here as FI for the full observation record) \cite{Komorowski2025}.

\subsection*{S1.2 Parallel independent receptors}

If we observe $N$ independent receptors in parallel (conditionally independent given $c$), their likelihood factorizes,
\begin{equation}
P(\bm Y\,|\,c)=\prod_{i=1}^N P(Y_i\,|\,c),
\end{equation}
so the FI adds,
\begin{equation}
\mathcal I_N(c)=N\,\mathcal I_1(c).
\end{equation}
Substituting into Eq.~\eqref{eq:SI_cap} gives
\begin{equation}
C_N^* = C_A^* + \frac{1}{2}\log_2 N,
\label{eq:SI_parallel}
\end{equation}
which is the $\tfrac12\log_2 N$ scaling quoted in the main text.

\subsection*{S1.3 Interpreting negative $C_A^*$ in plots}
In the main text figures we plot $C_A^*$ computed from Eq.~\eqref{eq:SI_cap} for a single receptor record. For short integration times (or weak sensitivity), the Fisher arc-length $V=\int\sqrt{\mathcal I(c)}\,dc$ can be smaller than $\sqrt{2\pi e}$, yielding $C_A^*<0$. This does not imply negative mutual information; it reflects the normalization of the asymptotic approximation and should be read as ``effectively zero bits'' at the single-receptor level under continuous sensing and continuous optimal input distribution \cite{Jetka2018,Komorowski2025}.

\section*{S2. Fisher information for a single two-state receptor (BP occupancy averaging)}
In the Berg--Purcell (BP) framework, the downstream circuit estimates concentration from the time-averaged occupancy
\begin{equation}
Y\equiv \frac{T_B}{T}=\frac{1}{T}\int_0^T n(t)\,dt,
\end{equation}
where $n(t)\in\{0,1\}$ indicates whether the receptor is bound.
For a stationary two-state Markov receptor, in the long-time limit $T\gg \tau_c$ the variance of the time average is
\begin{equation}
\mathrm{Var}(Y)\simeq \frac{2\,p(1-p)\,\tau_c}{T},
\end{equation}
with correlation time $\tau_c=(k_+c+k_-)^{-1}$. Error propagation through $p(c)=c/(c+K_d)$ (with $dp/dc=p(1-p)/c$) yields
\begin{equation}
\left(\frac{\delta c}{c}\right)^2
\equiv\frac{\mathrm{Var}(\hat c)}{c^2}
\simeq \frac{\mathrm{Var}(Y)}{[p(1-p)]^2}
=\frac{2\tau_c}{p(1-p)T}
=\frac{2}{k_-T}\,\frac{1}{p(c)}.
\label{eq:SI_bp}
\end{equation}
Using $p(c)=c/(c+K_d)$, Eq.~\eqref{eq:SI_bp} is equivalent to
\begin{equation}
\mathrm{Var}(\hat c)\simeq \frac{2}{k_-T}\,c(c+K_d).
\end{equation}
Inverting gives an effective FI (equivalently, the Cram\'er--Rao bound for an efficient estimator based on $Y$),
\begin{equation}
\mathcal I_0(c)\simeq \frac{1}{\mathrm{Var}(\hat c)}\simeq \frac{k_-T}{2}\,\frac{1}{c(c+K_d)}.
\label{eq:SI_FI_ideal}
\end{equation}

\section*{S3. Capacity for ideal timing: analytic integral}

Substituting Eq.~\eqref{eq:SI_FI_ideal} into Eq.~\eqref{eq:SI_cap} requires evaluating
\begin{equation}
V_0(c_{\max})\equiv \int_{0}^{c_{\max}}\sqrt{\mathcal I_0(c)}\,dc
=\sqrt{\frac{k_-T}{2}}\int_0^{c_{\max}}\frac{dc}{\sqrt{c(c+K_d)}}.
\end{equation}
Using the substitution $c=K_d\sinh^2 u$ gives $dc=2K_d\sinh u\cosh u\,du$ and
\begin{equation}
\sqrt{c(c+K_d)}=\sqrt{K_d\sinh^2u\cdot K_d\cosh^2u}=K_d\sinh u\cosh u,
\end{equation}
so the integral collapses to
\begin{equation}
\int_0^{c_{\max}}\frac{dc}{\sqrt{c(c+K_d)}}=2\,u_{\max}=2\,\mathrm{asinh}\!\left(\sqrt{\frac{c_{\max}}{K_d}}\right).
\end{equation}
Using $\mathrm{asinh}(z)=\ln(z+\sqrt{1+z^2})$ yields
\begin{equation}
V_0(c_{\max})= 2\sqrt{\frac{k_-T}{2}}\,\ln\!\left(\sqrt{1+\frac{c_{\max}}{K_d}}+\sqrt{\frac{c_{\max}}{K_d}}\right).
\label{eq:SI_V0}
\end{equation}
Finally,
\begin{equation}
C_N^* = \log_2\!\left[\frac{2}{\sqrt{2\pi e}}\,\ln\!\left(\sqrt{1+\frac{c_{\max}}{K_d}}+\sqrt{\frac{c_{\max}}{K_d}}\right)\right]
+\frac{1}{2}\log_2\!\left(\frac{k_-TN}{2}\right),
\label{eq:SI_C_main}
\end{equation}
which matches the closed-form capacity expression used in the main text.

\subsection*{S3.1 Large-dynamic-range asymptotics and the $\log\log$ divergence}

For $c_{\max}\gg K_d$, the logarithm behaves as
\begin{equation}
\ln\!\left(\sqrt{1+\frac{c_{\max}}{K_d}}+\sqrt{\frac{c_{\max}}{K_d}}\right)
=\ln\!\left(2\sqrt{\frac{c_{\max}}{K_d}}\right)+\mathcal O\!\left(\frac{K_d}{c_{\max}}\right).
\end{equation}
Therefore the capacity grows as
\begin{equation}
C_N^* \sim \log_2\ln c_{\max} + \frac{1}{2}\log_2(k_-TN) + \text{const.}
\end{equation}
The growth is unbounded but extremely slow, reflecting the fact that $\mathcal I(c)\propto c^{-2}$ at large $c$ implies constant precision in $\ln c$ even at arbitrarily large concentrations.

\section*{S4. Noisy clock model: derivation of the clock-imprecise FI and capacity formulas}

\subsection*{S4.1 What ``random $T$'' means here}

As discussed in the main paper, the idealized occupancy readout implicitly assumes that the elapsed observation time is known exactly. In a biochemical implementation, however, the downstream circuit must keep time with finite precision. Operationally, one can think of the receptor-bound time $T_B$ being integrated over a true window of intended duration $T$, while the downstream circuit must normalize by an imperfect clock variable $T_v$.
Thus $T_v$ is not a stochastic stopping time for the receptor trajectory in this model; it represents an independent duration-readout error in the normalization of an occupancy signal accumulated over the intended window $T$.
We take $T_v$ to be independent of receptor switching with
\begin{equation}
\langle T_v\rangle=T,\qquad \mathrm{Var}(T_v)=\sigma_T^2.
\end{equation}
The measurable occupancy-like readout is then
\begin{equation}
Y\equiv \frac{T_B}{T_v}.
\end{equation}
When $T_v=T$ deterministically, $Y$ reduces to the usual occupancy fraction $T_B/T$.

\subsection*{S4.2 Variance of $Y$ and induced concentration error}

For long $T$, the BP occupancy-based estimate gives a concentration error
\begin{equation}
\frac{\mathrm{Var}(\hat c)}{c^2}\simeq \frac{2}{k_-T}\,\frac{1}{p(c)}.
\end{equation}
Using the relation between $p$ and $c$,
\begin{equation}
\frac{d\ln c}{dp}=\frac{1}{p(1-p)},
\qquad\text{equivalently}\qquad
\frac{\delta c}{c}\simeq \frac{\delta p}{p(1-p)},
\end{equation}
and approximating $Y\approx p$ (so $\delta p\approx \delta Y$), we obtain
\begin{equation}
\mathrm{Var}(Y)\Big|_{\rm sensing}\simeq \mathrm{Var}(p) \simeq \frac{2}{k_-T}\,p(1-p)^2.
\label{eq:SI_varY_sense}
\end{equation}

Clock uncertainty enters because $Y=T_B/T_v$ depends on the denominator.
Linearizing in the clock error $\delta T_v=T_v-T$ with $\langle\delta T_v\rangle=0$ gives
\begin{equation}
Y=\frac{T_B}{T+\delta T_v}\approx \frac{T_B}{T}\left(1-\frac{\delta T_v}{T}\right)
= p\left(1-\frac{\delta T_v}{T}\right),
\end{equation}
so
\begin{equation}
\mathrm{Var}(Y)\Big|_{\rm clock}\simeq p^2\,\frac{\sigma_T^2}{T^2}.
\label{eq:SI_varY_clock}
\end{equation}
Assuming independence between receptor noise and clock noise, the two variances add:
\begin{equation}
\mathrm{Var}(Y)\simeq \frac{2}{k_-T}\,p(1-p)^2 + p^2\,\frac{\sigma_T^2}{T^2}.
\end{equation}
Propagating this to concentration using $\delta c/c\simeq \delta Y/[p(1-p)]$ gives
\begin{equation}
\left(\frac{\delta c}{c}\right)^2
\simeq
\underbrace{\frac{2}{k_-T}\,\frac{1}{p(c)}}_{\text{sensing noise}}
+\underbrace{\frac{\sigma_T^2}{T^2}\left(\frac{1}{1-p(c)}\right)^2}_{\text{clock noise}},
\label{eq:SI_relerr_clock}
\end{equation}
which is the clock-imprecise relative-error formula used in the main text.

\subsection*{S4.3 From variance to FI in the clock-noisy case}

Equation~\eqref{eq:SI_relerr_clock} provides an approximate lower bound on the achievable variance of any unbiased concentration estimator that relies on a noisy clock.
As in Sec.~S2, this implies an approximation of FI.
Writing the result in terms of FI gives
\begin{equation}
\mathcal I(c)^{-1}\simeq
\frac{2}{k_-T}\,c(c+K_d)
+\frac{\sigma_T^2}{T^2}\,\frac{c^2(c+K_d)^2}{K_d^2}.
\label{eq:SI_FI_clock}
\end{equation}
This matches the clock-imprecise FI used in the main text.
Notably, the second term forces $\mathcal I(c)\sim c^{-4}$ as $c\to\infty$, which guarantees a finite capacity as $c_{\max}\to\infty$.

\subsection*{S4.4 Closed-form evaluation of $\int\sqrt{\mathcal I(c)}dc$}

Define the dimensionless concentration $x\equiv c/K_d$, the dimensionless clock noise $s^2\equiv(\sigma_T/T)^2$, and $A\equiv 2/(k_-T)$.
Then Eq.~\eqref{eq:SI_FI_clock} becomes
\begin{equation}
\mathcal I(c)^{-1}=K_d^2\,x(1+x)\big[A+s^2 x(1+x)\big].
\end{equation}
Therefore
\begin{equation}
\int_{c_{\min}}^{c_{\max}}\sqrt{\mathcal I(c)}\,dc
=\sqrt{\frac{k_-T}{2}}\int_{x_{\min}}^{x_{\max}}\frac{dx}{\sqrt{x(1+x)\big[1+\beta x(1+x)\big]}},
\qquad \beta\equiv \frac{s^2k_-T}{2}.
\label{eq:SI_V_clock_start}
\end{equation}

Introduce an angle $\phi$ by the substitution
\begin{equation}
\sin\phi\equiv \frac{2\sqrt{x(1+x)}}{1+2x}.
\label{eq:SI_phi_def}
\end{equation}
This choice is equivalent to $x=(1-\cos\phi)/(2\cos\phi)$, which implies
\begin{equation}
\sqrt{x(1+x)}=\frac{1}{2}\tan\phi,
\qquad
1+2x=\sec\phi,
\qquad
dx=\frac{1}{2}\tan\phi\,\sec\phi\,d\phi.
\end{equation}
Substituting into Eq.~\eqref{eq:SI_V_clock_start} yields a standard elliptic integral:
\begin{align}
\int\frac{dx}{\sqrt{x(1+x)\big[1+\beta x(1+x)\big]}}
&=\int \frac{d\phi}{\sqrt{\cos^2\phi + (\beta/4)\sin^2\phi}}\\
&=\int \frac{d\phi}{\sqrt{1-\mu\sin^2\phi}},
\qquad
\mu\equiv 1-\frac{\beta}{4}=1-\frac{s^2k_-T}{8}.
\end{align}
Therefore
\begin{equation}
\int_{c_{\min}}^{c_{\max}}\sqrt{\mathcal I(c)}\,dc
=\sqrt{\frac{k_-T}{2}}\Big[\mathrm F(\phi_{\max}\,|\,\mu)-\mathrm F(\phi_{\min}\,|\,\mu)\Big],
\label{eq:SI_V_clock_closed}
\end{equation}
where $\mathrm F(\phi|\mu)=\int_0^{\phi} d\theta/\sqrt{1-\mu\sin^2\theta}$ is the incomplete elliptic integral of the first kind and $\phi(c)$ is defined by Eq.~\eqref{eq:SI_phi_def} with $x=c/K_d$.
Plugging Eq.~\eqref{eq:SI_V_clock_closed} into Eq.~\eqref{eq:SI_cap} gives the clock-imprecise capacity formula reported in the main text.

\end{document}